# Embedded transfer RNA Genes

**Zhumur Ghosh [(a)], Smarajit Das [(a)], Jayprokas Chakrabarti [(a, b,*)]**
**Bibekanand Mallick [(a)] and Satyabrata Sahoo [(a)]**

**(a) Computational Biology Group (CBG)**
Theory Department
Indian Association for the Cultivation of Science
Calcutta 700032 India

**(b) Biogyan**
BF 286, Salt Lake
Calcutta 700064 India

*** Author for correspondence**
Telephone: +91-33-24734971, ext. 281 (Off.)
Fax: +91-33-24732805
E-mail: tpjc@iacs.res.in

**Abstract:** In euryarchaeal methanogen *M.kandleri* and in Nanoarchaea *N. equitans* some of the missing tRNA genes are embedded in others. We argue from bioinformatic evidence that position specific intron splicing is the key behind co-location of these tRNA genes.

**Introduction:**

In our recent work[1] we analysed cytoplasmic tRNA genes ( tDNA ) of 22 species of 12 orders of three phyla of archaea. We looked for the identity elements for aminoacylation. During this investigation we found some tDNAs missing in euryarchaea and nanoarchaea. We observed later that some of these missing tDNAs lie embedded in other tDNAs. In this communication we argue that bioinformatic evidence points towards intron splicing at alternate positions in these embedded tDNAs. One composite tDNA gives rise to two different tRNAs.

The single-stranded primary tRNA nucleotide chain folds back onto itself to form the cloverleaf secondary structure. This structure has: (i) Acceptor or A arm: In this the 5/ and 3/ ends of tRNA are base-paired into a stem of 7 bp. (ii) DHU or D arm: Structurally a stem-loop, D-arm frequently contains the modified base dihydrouracil. (iii) Anticodon or AC arm, made of a stem and a loop containing the anticodon. At 5/ end of this anticodon-loop is a pyrimidine base at 32, followed by an invariant U at 33. The anticodon triplet is located at 34, 35, and 36 in the exposed loop region. (iv) An Extra Arm or V arm: This arm is not always present. It is of variable length and largely responsible for the variation in lengths of tRNAs. tRNA classification into types I and II depend on length of V-arm . (v) T-Ψ-C arm or T arm: This has conserved sequence of three ribonucleotides: ribothymidine, pseudouridine and cytosine. T arm has stem-loop structure and (vi) tRNA terminates with CCA at 3/ end. tDNA may or may not have CCA; If absent ,it is added during tRNA maturation.

Introns were found in several archaeal tDNAs between tRNA-nucleotide-positions 37 and 38, located in AC–loop[2]. These are the canonical introns (CI). Archaea, an intermediate between Eukarya and Bacteria, have tRNAs that share many similarities with either or both these domains[3]. Archaeal tDNAs harbour introns at various locations other than the canonical position of tRNA. These are the noncanonical introns (NCI).

Although, noncanonical introns in tDNAs were observed in 1987[4], bioinformatic identification of tDNAs harbouring these continues to be a challenge. As we developed our algorithm to circumvent some of the difficulties, we found evidence that tRNA genes overlap in archaea through introns, canonical and noncanonical. Earlier, in mitochondrial tRNAs, overlaps of between one to six nucleotides have been reported. $tRNA^{Tyr}$ and $tRNA^{Cys}$ genes in human mitochondrial genome, for instance[5] , overlap with one another by one nucleotide ( the last base of $tRNA^{Cys}$ and discriminator base of $tRNA^{Tyr}$ ). But tDNA-overlaps in archaea is an altogether new phenomena. In euryarchaeal methanogen *Methanopyrus kandleri*[6] AV19 (NC_003551) and in nanoarchaea *Nanoarchaeum equitans* Kin4-M [7] (NC_0005213) we find entire tDNAs embedded within one another.

Introns are present most frequently at the canonical position 37/38 in AC-loop. Apart from these, introns are also located in AC-arm, V-arm, D-arm and T-arm as well in A-arm[1]. The exact location of archaeal introns is obtained by looking for the presence of the bulge-helix-bulge (BHB) motif[8]. Archaeal splicing machinery cleaves introns at variable positions in tDNAs within the BHB motif[9]. In archaea, the tRNA endonuclease plays a key role in the removal of the intron from pre-tRNAs[10]. Hence, splicing of introns is a RNA-protein interaction which requires mutual recognition of two complementary tertiary structures.

## Methodology

The tRNA search programs like tRNAscan-SE and ARAGORN key on primary sequence patterns and/or secondary structures specific to tRNAs. A few loop-holes exist in these algorithms. It is the inability of these existing routines to identify tRNA genes if it harbours noncanonical introns in them. Some tRNAs are misidentified; some are missed out. We developed a computational approach to search for tDNAs that have noncanonical introns. With this algorithm we identified some non-annotated tDNAs. About one thousand tRNA-genes from archaea were studied for this purpose. From this database of 1000 tRNA-genes we fine-tuned the strategy to locate non-canonical introns. The salient features were :(i) sequences were considered that gave rise to the regular cloverleaf secondary structure. (ii ) conserved elements : T8 (except Y8 in *M. kandleri*), G18, R19, R53, T44, Y55, and A58 were considered as conserved bases for all archaeal tRNA . Further there were tRNA-specific conserved or identity elements[1] of archaea. (iii) the constraints of lengths of stems of regular tRNA A-arm, D-arm, AC-arm and T-arm were 7, 4, 5 and 5 bp respectively. In few cases the constraints on lengths of D-arm and AC-arm were relaxed. (iv)Promoter sequence ahead of the 5$'$-end looked for. ( v ) Base positions optionally occupied in D-loop were 17, 17a, 20a and 20b. (vi) Extra arm or V-arm was considered for tRNAs. The constraint on length of V-arm: less than 21 bases (vi) Noncanonical introns were considered at any position. The introns constrained to harbour the Bulge-Helix-Bulge (BHB) secondary structure for splicing out during tRNA maturation. The minimum length of introns allowed was 6 bases.

**Results and Conclusions:**

**tRNA$^{Gly}$ / tRNA$^{eMet}$ Embedded Genes of *M. kandleri***

This is our first example of embedded tRNA genes. In fig 1 we illustrate this embedding of two tDNAs.

tRNA$^{Gly}$(CCC) gene remained unidentified in *M. kandleri*. Note this gene is present in other archaea . Using our algorithm we identify it between c382165 and 382053. The sequence is presented is figure 1. This glycine tRNA has the important bases A73, C35 and C36 necessary for aminoacylation by AARS (aminoacyl tRNA synthetase)[11]. It has the conserved bases and base-pairs of other archaeal glycine tRNAs. In this tDNA, presumably[12], the 15 base long intron located at 32/33 is processed before splicing of second intron, 21 bases long, at 37/38. It has consensus BHB motif of type $h_e bh' bh' L$ (shown in figure 2). This sequential removal of introns implies that there is enough plasticity of tRNA molecule within the whole AC-stem and loop to allow major rearrangements between two successive splicing process .

One of the elongator methionine tRNA gene lies exactly embedded in this range. This eMet tRNA gene has all the important features of archaeal elongator methionine tRNA. C34, A35 and U36 are the identity elements in addition to the discriminator base A73 in this tRNA. It has a canonical intron of length 36. Part of the same BHB structure but this intron has a different splice-site marked in figure 2. The 3′ -splice-site for the canonical introns of the two embedded tRNA genes is the same .

**$tRNA^{Glu}$ / $tRNA^{eMet}$ Embedded Genes of *N. equitans***

This is the second example. Note in fig 1 we illustrate these embedded tRNA genes.

In *N. equitans* all the tRNA genes could not be located using tRNAscan-SE. Some of the missing ones were later found from the split-tRNA hypothesis[13,14] . We identify $tRNA^{Glu}$(CUC) in this genome lying between bases 327362 and 327500 of the genome. It has two introns, one canonical and one noncanonical. The canonical intron is 26 bases long. The noncanonical intron is located between bases 31 and 32 of AC-loop. The length of this noncanonical intron is 40 bases. The conserved bases and bps of archaeal $tRNA^{Glu}$ are consistent in this tRNA as well. U35 and C36 are identity elements for archaeal Glu tRNA as in *E. coli* [15,16]. C5:G68 could be another identity element[1] for archaeal $tRNA^{Glu}$ . All these identity elements are well present in this embedded $tRNA^{Glu}$ gene. The entire intronic structure has $h_eB[(h_1' L_1)$ $(h_2' L_2)$ $(h_3'$ b $h_3' L_3)]$ type BHB motif and has proper splice-sites ( figure 3).

The elongator methionine tRNA[17] gene also lies within this range. This tRNA has C34, A35 and U36 as the identity elements in addition to the discriminator base A73. These features are consistent with all other archaeal elongator methionine tRNA. It has a canonical intron of length 66. This also has the same BHB, albeit with different splicing position, marked in fig 3. The canonical introns of the embedded tRNA genes, once again, have the same 3′ - splice site.

In some of the primary transcripts of mitochondrial tRNA of animals, tRNAs are known to overlap by one to several bases [18]. In archaea we find tRNAs fully embedded in one other. The release of the entire versions of the two embedded tRNAs is assumed to occur . We believe one of the tRNAs is correctly processed in some transcripts, the other in other transcripts, potentially producing both complete transcripts. In these possibilities, the mode of recognition between the primary transcript and the processing enzyme(s) remains unclear. Presumably there exist sequence/structural patterns within the precursor tRNA, upstream or downstream, encoding this embedding. We are investigating features of pre-tRNA responsible for alternate endonucleolytic splicing of introns.

**Acknowledgements:**

We acknowledge useful discussions with Chanchal Dasgupta and Siddhartha Roy.

***M. kandleri***

tRNA$^{Gly}$ gene c(382165 .. 382053)

GCCGGGGCAGCTTAGCCTGGTAGAGCGCGGGGC**TCATAGG**
**GCCCGATG**ACCCC**GAGGGGTGACCGGCCTG GGAT**ACCCCG
AGGTCCCGGGTTCAAATCCCGGCCCCGGCACCA

tRNA$^{eMet}$ gene c(382165 .. 382053)

GCCGGGGCAGCTTAGCCTGGTAGAGCGCGGGGCTCATA**GG**
**GCCCGATGACCCCGAGGGGTGACCGGCCTGGGAT**ACCCCG
AGGTCCCGGGTTCAAATCCCGGCCCCGGCACCA

***N. equitans***

tRNA$^{Glu}$ gene (327362..327500)

GCCGCCGTAG CTCAGCGGTCAGAGCGCCCGG**CTCATAGCA**
**TGGGCTATGAGCTCTGACCCGAAAGGGGATG**ATCTCG**GGG**
**GCTCTTATGCCCCCTCGTGAGAA**ACCGGGA GGTCGCGGGT
TCGAATCCCG CCGGCGGCA

tRNA$^{eMet}$ gene (327362..327500)

GCCGCCGTAGCTCAGCGGTCAGAGCGCCCGGCTCATA**GCA**
**TGGGCTATGAGCTCTGACCCGAAAGGGGATGATCTCGGGG**
**GCTCTTATGCCCCCTCGTGAGAA**ACCGGGAGGTCGCGGGT
TCGAATCCCG CCGGCGGCA

Figure 1

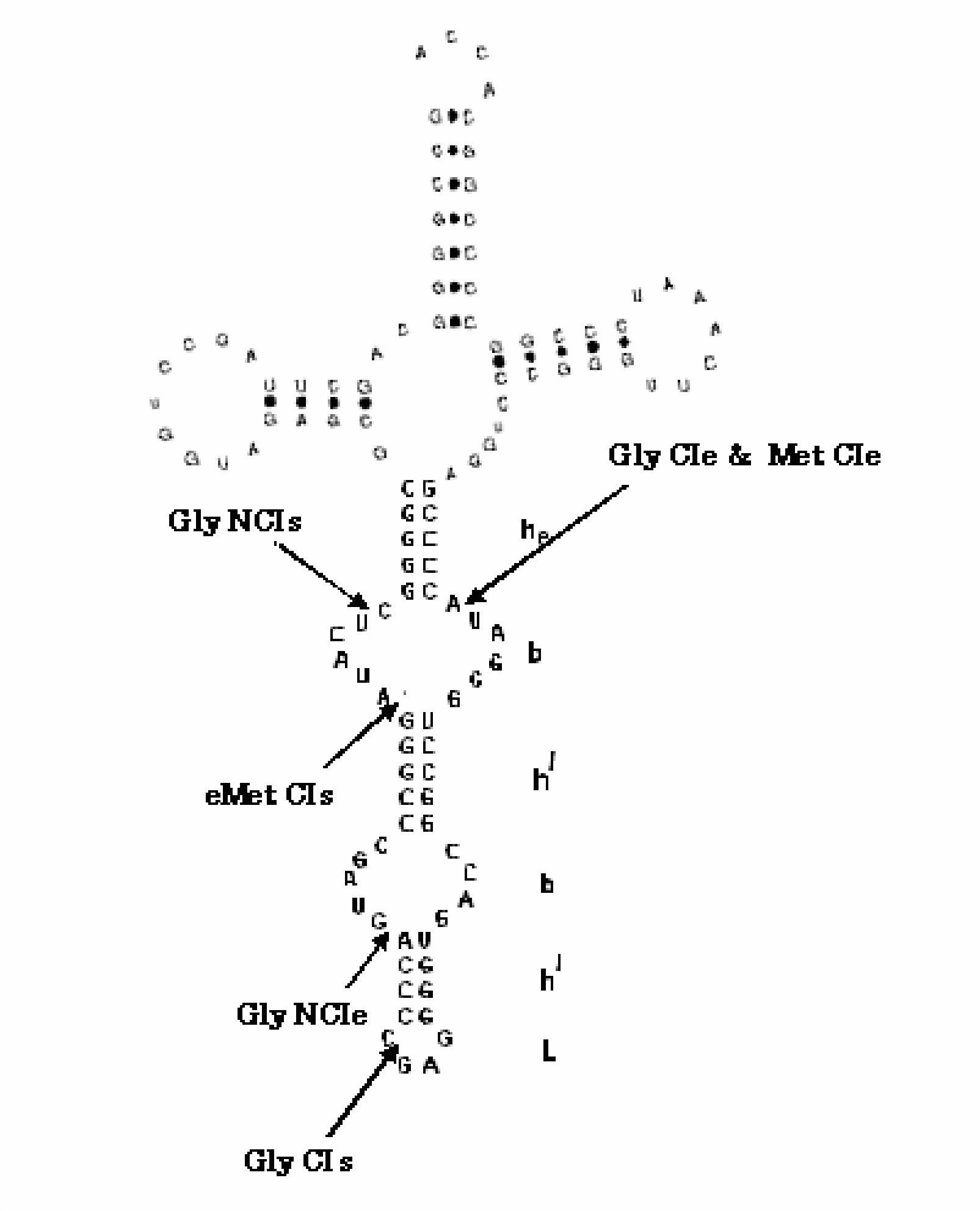

Figure 2

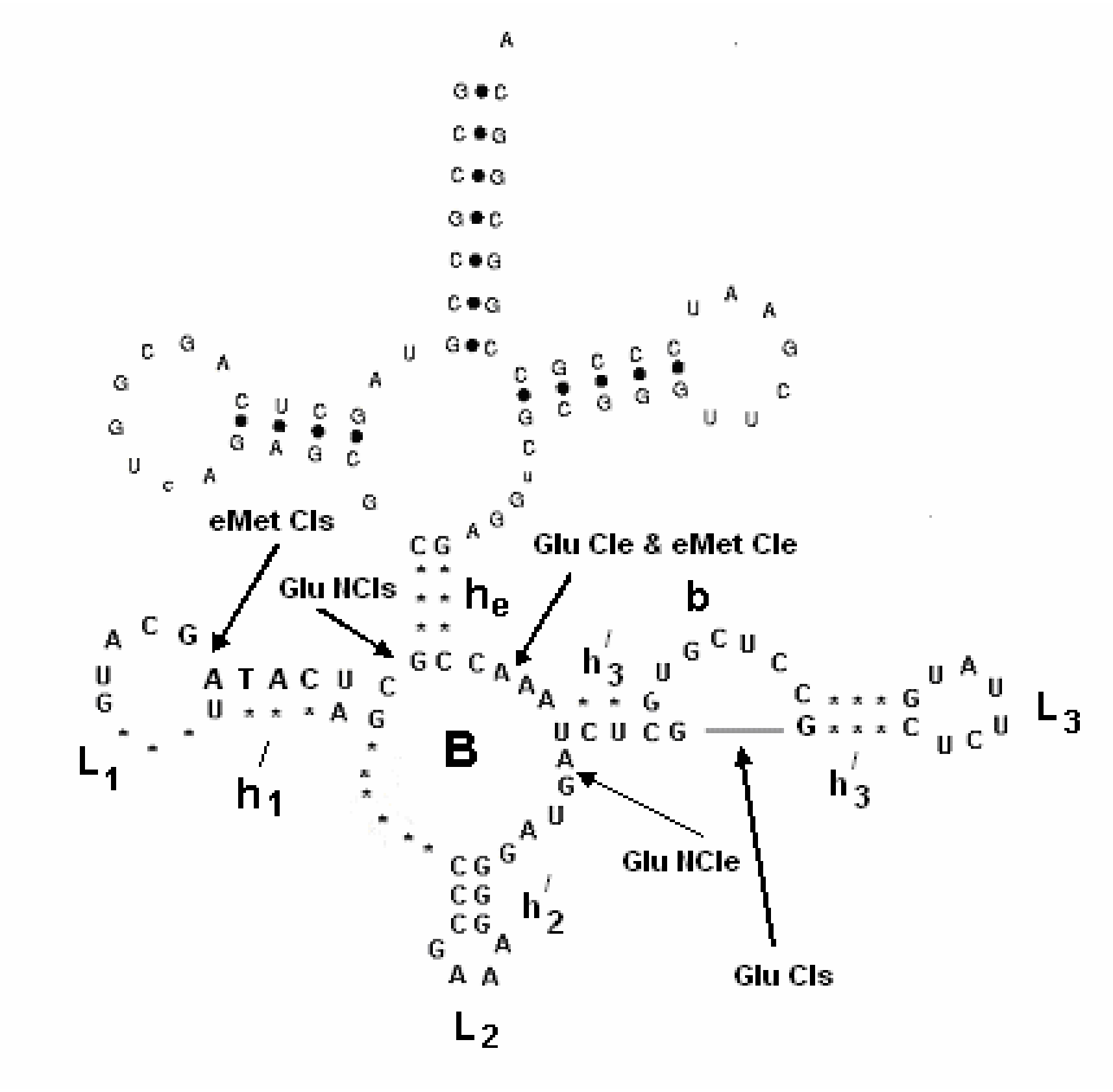

Figure 3

## Figure Legends

Figure 1: Sequences of the embedded tRNA genes.

Blue coloured region denotes noncanonical intron, brown canonical intron and black the exonic region.

Figure 2: Secondary structure of $tRNA^{Gly}$(CCC) / $tRNA^{eMet}$(CAT) along with the BHB of their introns of *M. kandleri*.

NCIs: Noncanonical intron start position; NCIe: Noncanonical intron end position;

CIs: Canonical intron start position; CIe: Canonical intron end position

→ This signifies splicing sites of the introns in pre-tRNAs

$h_e$: Exonic helix; $h^{/}$ : mixed helix (part of it is intronic and part of it is exonic)

b: bulge; L: loop

Figure 3: Secondary structure of $tRNA^{Glu}$(CTC) / $tRNA^{eMet}$(CAT) along with the BHB of their introns of *N. equitans*.

NCIs: Noncanonical intron start position; NCIe: Noncanonical intron end position;

CIs: Canonical intron start position; CIe: Canonical intron end position

→ This signifies splicing sites of the introns in pre-tRNAs

$h_e$: Exonic helix; $h^{/}$: mixed helix ($h_1^{/}$ : mixed helix in the $1^{st}$ branch; $h_2^{/}$ : mixed helix in the $2^{nd}$ branch; $h_3^{/}$ : mixed helix in the $3^{rd}$ branch )

L : loop ($L_1$: loop in the $1^{st}$ branch; $L_2$ : loop in the $2^{nd}$ branch; $L_3$ : loop in the $3^{rd}$ branch )